\documentclass[twocolumn,aps,prl,preprintnumbers,showpacs,nofootinbib]{revtex4}

\usepackage{color}
\usepackage{overpic}
\usepackage{subfigure}
\usepackage[english]{babel}
\usepackage{amsmath}
\usepackage{amssymb}
\usepackage{enumerate}
\usepackage{epsfig}
\usepackage{graphics,psfrag,rotating}
\usepackage{graphicx}
\usepackage{appendix}
\usepackage{dcolumn}
\usepackage{bm}
\usepackage{cleveref}
\usepackage{float} 
\usepackage[utf8]{inputenc}
\usepackage[english]{babel}
\usepackage{amsmath}
\usepackage{amsfonts}
\usepackage{amssymb}
\usepackage{epsfig}
\usepackage{graphics,psfrag,rotating}
\usepackage{graphicx}
\usepackage{dcolumn}
\usepackage{bm}
\usepackage{epstopdf}
\usepackage{color}
\usepackage[normalem]{ulem} 
\usepackage{epsfig}
\usepackage{graphics,psfrag,rotating}
\usepackage{graphicx}
\usepackage{dcolumn}
\usepackage{bm}
\usepackage{epstopdf}
\usepackage{color}
\usepackage[usenames,dvipsnames,svgnames]{xcolor}
\usepackage[colorlinks=true,
            linkcolor=red,
            urlcolor=gray,
            citecolor=blue]{hyperref}
  \usepackage{hyperref}
\newcommand{\lsim}   {\mathrel{\mathop{\kern 0pt \rlap
{\raise.2ex\hbox{$<$}}}
 \lower.9ex\hbox{\kern-.190em $\sim$}}}
\newcommand{\gsim}   {\mathrel{\mathop{\kern 0pt \rlap
{\raise.2ex\hbox{$>$}}}
\lower.9ex\hbox{\kern-.190em $\sim$}}}

\def\3nab{\tilde{\nabla}}

\def\hsp5{\hspace{5mm}}

\def\case#1/#2{\textstyle\frac{#1}{#2}}

\def\ber {\begin{eqnarray}}
\def\eer {\end{eqnarray}}
\def\bea {\begin{eqnarray}}
\def\eea {\end{eqnarray}}

\def\bc {\begin{center}}
\def\ec {\end{center}}
\def\case#1/#2{\frac{#1}{#2}}

\newcommand{\bw}{\begin{widetext}}
\newcommand{\ew}{\end{widetext}}

\newcommand{\be}{\begin{equation}}
\newcommand{\bse}{\begin{subequation}}
\newcommand{\ese}{\end{subequation}}
\newcommand{\ee}{\end{equation}}
\newcommand{\eei}{\end{eqnarray}\indent\indent}
\newcommand{\ba}{\begin{array}}
\newcommand{\ea}{\end{array}}
\newcommand{\bal}{\begin{eqnarray}}
\newcommand{\eal}{\end{eqnarray}}

\def\case#1/#2{\textstyle\frac{#1}{#2} }

\newcommand{\bver}{\begin{verbatim}}
\newcommand{\ever}{\end{verbatim}}

\begin{document}

\preprint{SLAC-PUB-16024}

\title{Reissner-Nordstr\"{o}m Black Holes in the Inverse Electrodynamics Model}
\thispagestyle{empty}

\author{
J.~A.~R.~Cembranos$\,^{(a,b)}$\footnote{Email: cembra [at] fis.ucm.es},
A.~de la Cruz-Dombriz$\,^{(b)}$\footnote{Email: dombriz [at] fis.ucm.es} and
J.~Jarillo$\,^{(c)}\footnote{Email: jjarillo [at] ucm.es}$
}
\affiliation{$^{(a)}$ SLAC National Accelerator Laboratory, 2575 Sand Hill Rd, Menlo Park, CA 94025, USA;}
\affiliation{$^{(b)}$ Departamento de F\'{\i}sica Te\'orica I, Universidad Complutense de Madrid, E-28040 Madrid, Spain;}
\affiliation{$^{(c)}$ Departamento de F\'{\i}sica At\'omica, Molecular y Nuclear, Universidad Complutense de Madrid, E-28040 Madrid, Spain.}

\date{\today}





%
%
%


\begin{abstract}
We study electric and magnetic monopoles in static, spherically symmetric and constant curvature geometries in the context of the inverse
electrodynamics model. We prove that this U(1) invariant Lagrangian density is able to support the standard metric of a
Reissner-Nordstr\"{o}m Black Hole, but with more complex thermodynamical properties than in the standard case.
By employing the Euclidean Action approach we perform a complete analysis of its phase space depending on the sign and
singularities of the 
heat capacity and the Helmholtz free energy.
\end{abstract}

\maketitle


\section{Introduction}
\label{sec:Intro}

In General Relativity coupled with the usual U(1) invariant Electrodynamics theory, the Reissner-Nordstr\"om black-hole (BH) solution arises, corresponding to a massive, charged, non-rotating and spherically symmetric body \cite{Reissner, Nordstrom}. This kind of solution has been widely studied in the last decades  ({\it c.f.} Refs.~\cite{Peca:1998cs, Barnich:2007uu, Jardim:2012se}). Nevertheless the divergence of self-energy of point charges (like electrons) in the standard Electrodynamics theory has suggested that
modified Electrodynamics theories might be required in order to circumvent this shortcoming.
Nonlinear models have also been studied from the point of view of  effective Lagrangians which attempt to describe Quantum Electrodynamics \cite{QED}.
Some important examples of these kinds of theories are the Born-Infeld \cite{Born:1934gh, BI_others} and the Euler-Heisenberg models \cite{Heisenberg:1935qt, Dunne:2012vv,Fradkin:1985qd, Tseytlin:1997csa, Brecher:1998tv, Euler_others}. Following this line of reasoning, in the last years, different works have studied modified Electrodynamics models coupled with gravity \cite{Other_NonLinear}. In particular models providing static and spherically symmetric solutions for electrostatic spherically symmetric fields have drawn remarkable attention ({\it c.f.} \cite{DiazAlonso:2009ak} and references therein).

On the other hand, the study of the thermodynamics properties of BH solutions began in the 1970's with the attainment of the four laws of BHs dynamics \cite{Bardeen:1973gs}. These mechanics laws seem very similar to the four laws of Thermodynamics, where the BH mass, the area of the horizon and the surface gravity play analogous roles to the energy, the entropy and the temperature, respectively. One approach in order to compute the thermodynamical properties of a BH solution is the Euclidean Action Method \cite{Gibbons:1976ue, Hawking:1978jz}. 
%
%
%
The Euclidean approach exhibits some difficulties when is applied to General Relativity. Except in special cases it is generally impossible to represent an analytic spacetime as a Lorentzian section of a four-complex-dimensional manifold with a complex metric which possesses a Euclidean section. Therefore there is not a general prescription for analytically {\it continuing} Lorentzian signature metrics to Riemannian metrics. However, in static metrics on which we shall focus,
the aforementioned {\it continuation} procedure can be done. 
Nevertheless, even if possible to be performed, there are not any theorems guaranteeing the analyticity of the obtained quantities (for further details, {\it c.f.} Ref.~\cite{Wald:1984rg}, \cite{Hawking:1974rv}).

The paper is organized as follows: in Section \ref{sec:GR_model} we introduce the Inverse Electrodynamics Model (IEM) and the
 static, spherically symmetric solutions supported therein by electric and magnetic monopoles. In Section \ref{sec:termo} we apply the Euclidean Method in order to distinguish the  different thermodynamics phases of the solutions, defined in terms of their stability, and we shall compare the phase diagrams with the standard electrodynamics model counterparts. The appearance of a new thermodynamical phase, absent in the standard case, shall be extensively discussed. In Section \ref{sec:termo_speed} we then perform a classification of the BH configurations depending on the phase transitions that they present. Finally, in Section \ref{sec:conclusions} we summarize the main results and conclusions of the paper.

Unless otherwise specified, Planck units, $(G=c=k_B =\hbar=4\pi \varepsilon_0 =1)$ will be used throughout this paper, Greek indices run from 0 to 3. The symbol $\nabla$ denotes the standard covariant derivative and the signature $+,-,-,-$ is used.

\section{Inverse Electrodynamics Model}
\label{sec:GR_model}

In this section, we shall show the static and spherically symmetric solutions for the IEM in General Relativity.
Thus the action is given by
\begin{eqnarray}
S=S_g+S_{{\rm U(1)}}\,,
\label{descomposicion_accion}
\end{eqnarray}
where $S_g$ and $S_{{\rm U(1)}}$ denote  the gravitational and matter terms of the action, respectively. The usual gravitational action term takes the form
\begin{eqnarray}
S_g = \frac{1}{16 \pi} \int {\rm d}^4x \sqrt{\left| g \right|} \left( R-2 \Lambda \right) \,,
\end{eqnarray}
being $g$ the determinant of the metric $g_{\mu \nu}$, 
$R$ the scalar of curvature and $\Lambda$ a cosmological constant.

On the other hand, we assume that the matter term of the action, $S_{{\rm U(1)}}$, is given by the IEM Lagrangian density $\mathcal{L} (X,Y)$, namely,
\begin{eqnarray}
\mathcal{L} (X,Y) = \frac{1}{8\pi} X \left[ 1 - \eta \left( \frac{Y}{X} \right)^2\right]\,,
\label{phimodelo}
\end{eqnarray}
which is a function of the Maxwell invariants $X$ and $Y$, defined as
\begin{eqnarray}
X \equiv -\frac{1}{2} F_{\mu \nu} F^{\mu \nu}\,, \,\,\,
Y \equiv -\frac{1}{2} F_{\mu \nu} F^{*\,\mu \nu} \label{XY_paso1} \,,
\end{eqnarray}
being $F_{\mu \nu}=\partial_\mu A_\nu - \partial_\nu A_\mu$ the usual electromagnetic tensor and $F^*_{\,\mu \nu} \equiv \frac{1}{2}\sqrt{\left| g\right|}\epsilon_{\mu \nu \alpha \beta}F^{\alpha \beta}$, with $\epsilon_{\mu \nu \alpha \beta}$ the Levi-Civita symbol. In terms of the Lagrangian density $\mathcal{L} (X,Y)$, the matter term  of the action  \eqref{descomposicion_accion} takes the form
\begin{eqnarray}
S_{{\rm U(1)}} = - \int {\rm d}^4x \sqrt{\left|g \right|} \mathcal{L} (X,Y)\,.
\label{MatterAction}
\end{eqnarray}
This action is parity-invariant and can be interpreted as a perturbation of the standard Electrodynamics theory ($\mathcal{L}(X,Y) \sim X$) for a small enough value of the parameter $\eta$. Moreover, provided $F_{\mu \nu}$ represents an electric monopole with a null magnetic field, the standard Lagrangian and the standard point-like solutions are recovered as one might expect. Another interesting property of the IEM is its conformal invariance. In fact, the trace of the associated energy-momentum tensor  vanishes as in standard Electrodynamics, i.e.,
\begin{eqnarray}
T\equiv T^\mu_{\,\,\,\mu} =
g^{\mu\nu}T_{\mu\nu}&=&-\frac{2g^{\mu\nu}}{\sqrt{\mid g\mid}}\frac{\delta S_{U(1)}}{\delta g^{\mu\nu}}=0.
\label{Tmunu}
\end{eqnarray}
In this paper we restrict ourselves to the study of static and spherically symmetric solutions. Hence,  for the metric tensor let us consider the most general ansatz for static and spherically symmetric scenarios,
\begin{eqnarray}
\!\!\!\! {\rm d}s^2= \lambda(r){\rm d}t^2 - \frac{1}{\mu(r)}{\rm d}r^2-r^2 \left( {\rm d}\theta^2+ \sin ^2 \theta {\rm d}\phi^2\right)\,, \label{gtensor}
\end{eqnarray}
where the functions $\lambda (r)$ and $\mu (r)$ depend solely on $r$ in order to ensure staticity and spherical symmetry. Besides, with this metric \eqref{gtensor} we consider an ansatz for the electromagnetic tensor
\begin{eqnarray}
\!\!F_{01}=-F_{10}=E(r) \,, \,\,\, F_{23}=-F_{32}=-B(r)r^2 \sin \theta\,,
\label{tensor_electromagnetico}
\end{eqnarray}
being identically null the other components, and $E(r)$ and $B(r)$ functions on $r$. In Minkowski spacetime, where $\lambda(r)$ and $\mu(r)$ equal to $1$, \eqref{tensor_electromagnetico} is the electromagnetic tensor for radial electric and magnetic fields $E(r)$ and $B(r)$, respectively \cite{Jackson}. For this reason, we shall refer to these functions as ``electric'' and ``magnetic'' fields.

With the metric \eqref{gtensor}, 
the gauge invariants \eqref{XY_paso1} can be rewritten in terms of the electric and magnetic fields as follows
 \begin{eqnarray}
\!\!\!\!X =\frac{\mu(r)}{\lambda(r)} E(r)^2-B(r)^2 \,,\,\,\,
Y = 2 \sqrt{\frac{\mu(r)}{\lambda(r)}} E(r) \cdot B(r) \label{XY_paso2} \,.
\end{eqnarray}

By performing variations of the total action \eqref{descomposicion_accion} with respect to the metric tensor, we achieve the Einstein field equations in metric formalism,
\begin{eqnarray}
 R_{\mu\nu} - \frac{1}{2}R g_{\mu\nu}+\Lambda g_{\mu \nu} = 8\pi  T_{\mu\nu}   \,,
 \label{ec_campo}
\end{eqnarray}
where $ R_{\mu\nu} $  holds for the Ricci Tensor. 

Furthermore, by replacing the Lagrangian density \eqref{phimodelo}, and the metric and electromagnetic tensors \eqref{gtensor} and \eqref{tensor_electromagnetico} in the energy-momentum tensor definition \eqref{Tmunu}, the
non zero components of the latter tensor can be found. Together with the metric tensor above \eqref{gtensor}, these components enable the resolution of the field equations \eqref{ec_campo} yielding
\begin{eqnarray}
\lambda (r) = \mu (r)  \label{lmec0}\,.
\end{eqnarray}
With this expression, the gauge invariants \eqref{XY_paso2} can be  simplified  reading
$X = E(r)^2-B(r)^2$ and 
$Y=   2 E(r)\cdot B(r)$,  
i.e., the usual gauge invariants in standard Electrodynamics are recovered. Moreover, we can replace \eqref{lmec0} in the field equations  \eqref{ec_campo},
achieving the expressions
\begin{eqnarray}
    - r \lambda'(r)- \lambda(r)     +1  +\Lambda r^2= 8 \pi T^0_{\;\;0} (r) r^2  \,, \label{lmec1}  \\
     2  \lambda'(r)       + r \lambda''(r)  -2\Lambda r= -16 \pi T^2_{\;\;2} (r) r\,. \label{lmec2}
\end{eqnarray}
%
%
%
The general solution of the field equations system (\ref{lmec1})-(\ref{lmec2}) reads
\begin{eqnarray}
\lambda (r) = 1 -\frac{2M}{r} + \frac{2\varepsilon_{ex}(r)}{r} 
+\frac{1}{3} \Lambda r^2\,,
\label{lambda_general_vs_T00}
\end{eqnarray}
where $M$ is an integration constant, that can be identified as the BH mass and $\varepsilon_{ex} (r) \equiv 4 \pi \int _r^\infty x^2 T^0_{\;\;0} (x) {\rm d}x\,$, dubbed {\it external energy},
can be understood as the energy provided by the U(1) fields $E(r)$ and $B(r)$ outside a sphere of radius $r$ \cite{DiazAlonso:2009ak}.
%
%

Considering now $\mathcal{L}(X,Y)$ and its derivatives, the associated Maxwell's field equations, together with the Bianchi identities for the electromagnetic field, take the form
\begin{eqnarray}
	\nabla _\mu \left( \mathcal{L} _X F^{\mu \nu} + \mathcal{L} _Y F^{* \mu \nu} \right) = 0 \,,\,\,\,\, \nabla _\mu F^{*\mu \nu}=0 \,.
\label{MAX_N}
\end{eqnarray}
These generalized Maxwell's equations can be expressed for static and spherically symmetric solutions of the IEM with the electromagnetic tensor \eqref{tensor_electromagnetico} as
\begin{eqnarray}
r^2 B(r) = Q_t \,,
\label{campoB}
\end{eqnarray}
\begin{eqnarray}
&&r^2   \left[  1 + 4 \eta \left( \frac{E(r) B(r)}{E(r)^2-B(r)^2} \right)^2  \right] E(r) \nonumber \\
&&= 4 \eta  \frac{E(r) B(r)}{E(r)^2-B(r)^2} Q_t  + Q_c \,
\label{campoE2} \,,
\end{eqnarray}
with $Q_c$ and $Q_t$, i.e., the current and the topological charges respectively, acting as sources. 
It is easy to see that equation (\ref{campoE2})
possesses solutions for electric fields that decrease as $r^{-2}$. Thus, provided that we impose $E(r) = Q_e / r^2$, and using equation (\ref{campoB}),  we achieve an equation for this parameter $Q_e$
\begin{eqnarray}
Q_e \left[ 1 +4 \eta \left( \frac{ Q_e Q_t}{Q_e^2-Q_t^2}\right)^2\right]
= 4 \eta  \frac{Q_e Q_t^2}{Q_e^2 - Q_t^2 } +  Q_c\,.
\label{campoE3}
\end{eqnarray}
From this equation, one can obtain the parameter $Q_e$ as a function of  $\eta$ and the charges $Q_c$ and $Q_t$, and seeing that $Q_e$ coincides with the current charge $Q_c$ in standard Electrodynamics ($\eta=0$). The analytic expression of this parameter is not trivial, but for a small enough $\eta$, could be expressed as
\begin{eqnarray}
Q_e =  Q_c - 4 \eta \frac{Q_c Q_t^4}{\left( Q_c^2-Q_t^2\right)^2} +\mathcal O(\eta^2)\,, \label{desarrollo_Theta}
\end{eqnarray}
whereas if the topological charge is smaller than the current one, the expression reads
%
%
\begin{eqnarray}
Q_e = Q_c\left[1 - 4 \eta \left( \frac{Q_t}{Q_c} \right)^4 + \mathcal{O}  \left(  \frac{Q_t}{Q_c} \right)^6 \right]\,.
\end{eqnarray}
In the following,  instead of using as charges $\left\{ Q_c, Q_t \right\}$ we choose $\left\{Q_e, Q_m \right\}$ (being $Q_m \equiv Q_t$), denoted as ``electric'' and ``magnetic'' charges. This election has the important advantage that the electric and magnetic fields read directly as $E=Q_e/r^2$ and $B=Q_m /r^2$ and therefore the interpretation of the following results.

\begin{figure*}
	\centering
		\includegraphics[width=0.38\textwidth]{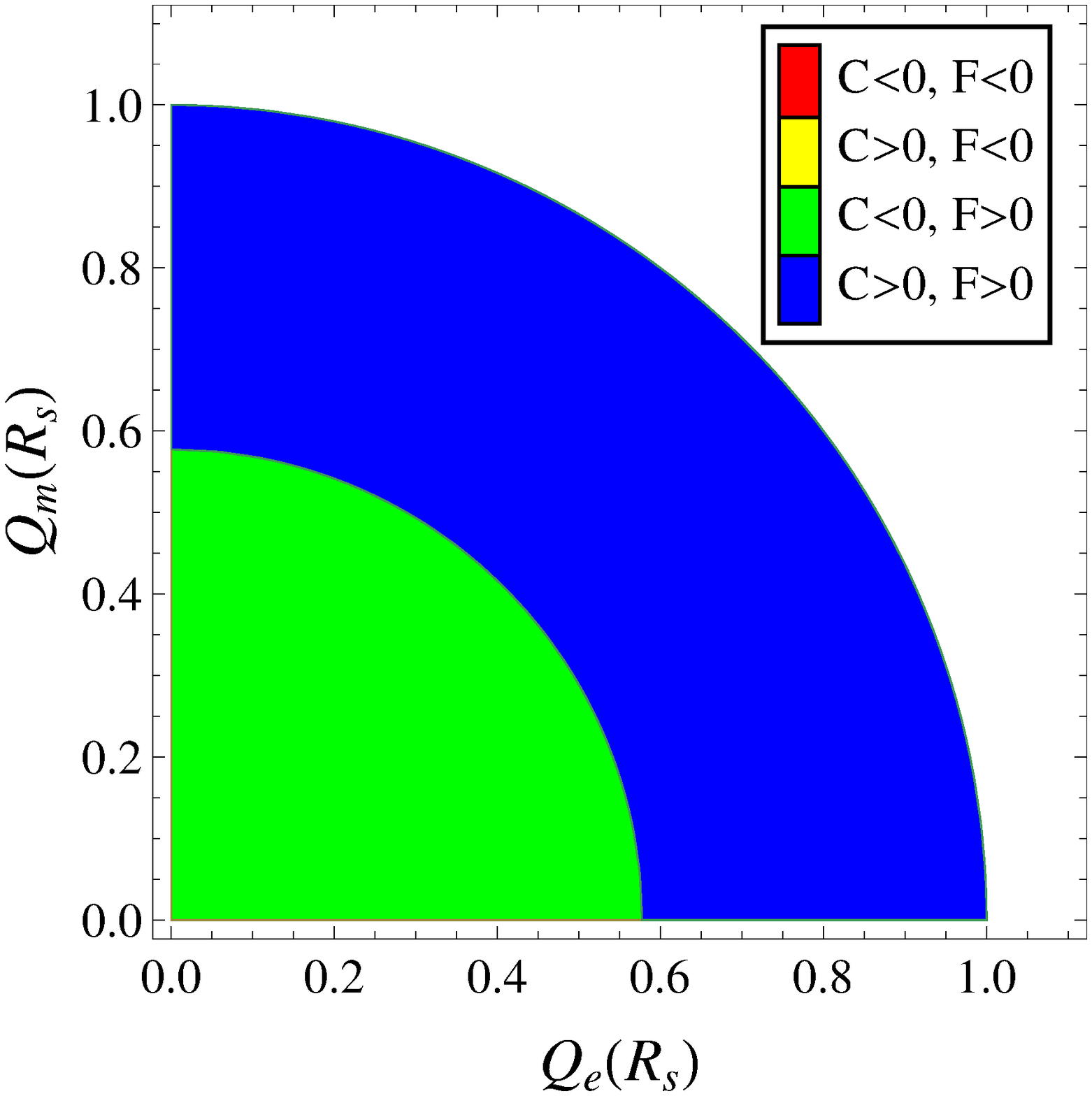}\,\,\,\,\,\,\,\,\,\,\,\,\,\,
		\includegraphics[width=0.38\textwidth]{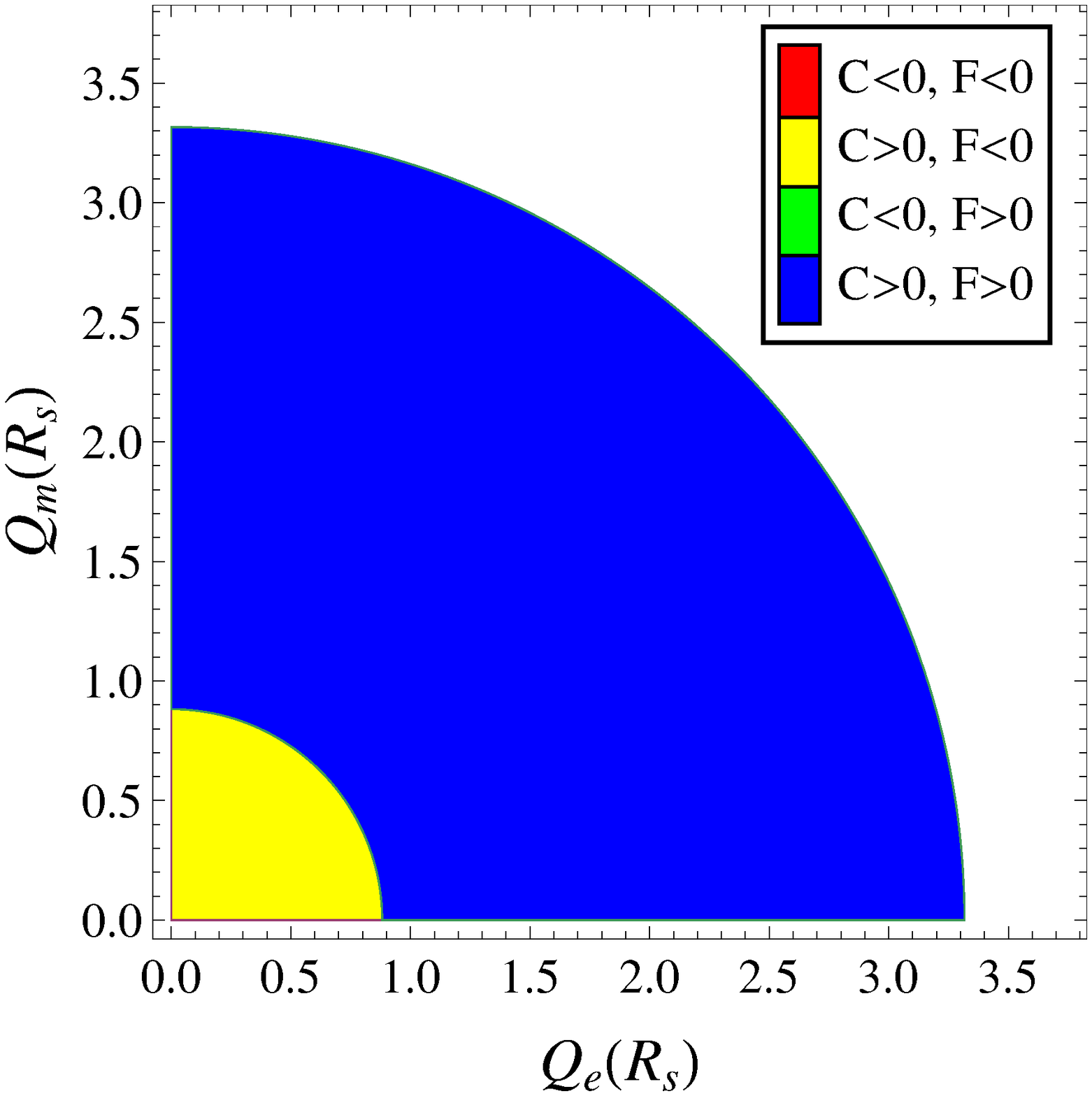}
		\caption{\footnotesize{
		Phase diagrams of BH solutions with $\eta=0$ (usual Electrodynamics Lagrangian) corresponding to $r_{h}=R_s$, in flat spacetime ($\Lambda=0$) (left panel) and AdS spacetime with $\Lambda=10 R_s^{-2}$ (right panel), being $R_s$ the Schwarzschild radius of an object with a solar mass, $R_s\simeq 10^{38} l_p$. The electric and magnetic charges are also expressed in $R_s$ Planck charges. 
In the flat spacetime case, two different phases exist: in blue both $C$ and $F$ are positive, while in green $C<0$ and $F>0$. For AdS scenario $\Lambda=10 R_s^{-2}$  there is a phase with $C>0$ and $F<0$ (yellow) and again a phase with both quantities positive. The phase with both $C$ and $F$ negative does not hold on for the usual Electrodynamics theory. Regions in white correspond to masses below the extremal BH mass. The diagram is represented solely for positive values of the charges; however,  under the reversal $Q_e \rightarrow - Q_e$ or $Q_m \rightarrow -Q_m$ the diagram would be completely  symmetric.}}
	\label{fig:termo_estandar}
\end{figure*}

After performing some intermediate calculations involving the determination of $T^{0}_{\;0}$ component, one can get an expression for $\varepsilon_{ex}(r)$ defined after
the equation (\ref{lambda_general_vs_T00}), 
and rewrite the {\it external energy} in this case in the form
%
%
\begin{eqnarray}
\varepsilon_{ex} (r)\,=\,\frac{Q_e^2 +Q_m^2}{2r} \left[  1 +4 \eta \left( \frac{Q_e Q_m}{Q_e^2- Q_m^2}\right)^2 \right]\,\equiv\,\frac{\mathcal{K}}{2r} .
\label{K2(Theta,Qm)}
\end{eqnarray}
where  the parameter $\mathcal{K}$ has been introduced for convenience.
One can realize that, as also occurs in the standard case ($\eta=0$) the external energy diverges at the origin, i.e., the total energy from the U(1) fields is divergent.
Furthermore, we can replace \eqref{K2(Theta,Qm)} in the expression \eqref{lambda_general_vs_T00},  so the metric parameter $\lambda(r)$ can be rewritten as
\begin{eqnarray}
\lambda(r)=1-\frac{2M}{r}+ \frac{\mathcal{K}}{r^2}+\frac{1}{3}\Lambda r^2\,,
\label{lambda_general}
\end{eqnarray}
The obtained metric  corresponds to a Reissner-Nordstr\"{o}m-like with a scalar curvature $R=4\Lambda$, and a modified charge term equal to $\mathcal{K}$ which in the standard case ($\eta=0$) provides the well-known sum of squares of charges $Q_e^2+Q_m^2$.
Once the metric parameter $\lambda(r)$ for the Reissner-Nordstr\"om-like solution has been obtained, the horizons structure 
can be determined. In order to obtain the radii of the horizons, one has to calculate the roots of $\lambda(r)$ or, equivalently, satisfying the condition
\begin{eqnarray}
M-\frac{r_h}{2}-\frac{1}{6} \Lambda r_h^2 = \varepsilon_{ex} (r_h)\,,
\label{M-horizontes-rubiera1}
\end{eqnarray}
whose solutions may provide in general one external (event) horizon and one internal horizon. Note that the external energy could be either positive or negative depending on the sign of $\mathcal{K}$. We are just interested in the anti-de Sitter (AdS) case $\Lambda>0$,
since otherwise $(\Lambda<0)$ some problems of normalization of the temporal Killing $\partial_t$ arise \cite{Gibbons:1977mu}.
Thus,  the value of the external horizon yields \cite{Sahay:2010tx}
\begin{eqnarray}
r_{h} &=& \frac{1}{2} \left( \sqrt{x} + \sqrt{-\frac{6}{\Lambda}-x+\frac{12 M}{\Lambda \sqrt{x}}} \,\right)   \,,
\label{horizontes}
\end{eqnarray}
with
\begin{eqnarray}
x&=&  \left( \frac{1+4\Lambda \mathcal{K}}{\Lambda}\right)  \sqrt[3]{\frac{2}{y}}+\frac{3}{\Lambda}\sqrt[3]{\frac{y}{32}}-\frac{2}{\Lambda}\,,
\label{x}
\end{eqnarray}
and
\begin{eqnarray}
\!\!\!\!\!\!\!y &=& 2+36\Lambda M^2 -24 \Lambda \mathcal{K} \nonumber \\
	&+&\sqrt{\left(  2+36 \Lambda M^2 -24 \Lambda \mathcal{K} \right)^2-4\left(  1+4\Lambda \mathcal{K} \right)^3}\,.
\label{y}
\end{eqnarray}
Then, using \eqref{M-horizontes-rubiera1} we can write the BH mass as a function of the external horizon radius $r_h$, the charge term $\mathcal{K}$ and the cosmological constant $\Lambda$, provided that at least one horizon is present, as
\begin{eqnarray}
M(r_h)=\frac{r_h}{2} \left(1+\frac{\mathcal{K}}{ r_h^2}+\frac{1}{3} \Lambda r_h^2 \right)\,.
\label{funcionM}
\end{eqnarray}
If we assume both $\mathcal{K}$ and $\Lambda$ positive (as in the standard AdS case), the function $M(r_h)$ has a minimum at $r_{h\,\,min} = \sqrt{ \left( \sqrt{1+4 \Lambda \mathcal{K}}-1\right)/2\Lambda}$. This means that provided the mass of the configuration is small enough, no horizon appears and then such configuration would not constitute a proper BH.
Hence, the condition to have at least one horizon and then have a truly BH solution $r_h^2  \ge r_{h\,\,min}^2$, can be summarized as
\begin{eqnarray}
r_h^2 \left( 1+ \Lambda r_h^2   \right) \ge \mathcal{K}  \,,
\label{condition_extremal}
\end{eqnarray}
where the inequality saturates for the extremal BH.
However, provided $\mathcal{K}$ takes negative values, which necessarily requires $\eta<0$, and $\Lambda$ is non negative, the range of values of $M(r_h)$ may entirely cover the interval  $\left[ 0, \infty \right)$. Then, in the latter scenario (${\mathcal K}<0$) it would be possible to host a BH solution with at least one horizon for an arbitrary positive - or even negative -mass value, unlike the standard Electrodynamics case, for which there is always a BH configuration which is extremal. Concerning the possibility of negative values for the parameter $M$, despite the fact that it is allowed by the analysis above, at least from two points of view, we must 
conclude that realistic physical configurations force $M$ to be positive (or null). First, at large distances from the black-hole configurations, expression (\ref{lambda_general}) 
must result in the well-known Schwarzschild-(Anti)-de Sitter limit or complementary, the Newtonian interpretation of the metric coefficients as gravitational potentials for a weak gravitational field. This requirement forces the parameter $M$ to be positive and to be interpreted as the total mass of the configuration. Secondly, the massive energy of the configuration that we shall introduce in Eqn. (\ref{Massive_energy}) proves that this energy correctly coincides with $M$. Since as a consequence of the attractive character of gravity, this energy is usually considered as positive, we are left with a supplementary reason to consider the parameter $M$ as positive for viable physical configurations.


Finally, in order to obtain the mass of the extremal BH in the IEM one must replace  the expression for $r_{h\,min}$ in Eq.~\eqref{funcionM}, yielding
\begin{eqnarray}
M_{extr}=\frac{\sqrt{2}}{6} \frac{\sqrt{1+4\Lambda \mathcal{K}} -1 +4\Lambda \mathcal{K}}{\sqrt{\Lambda} \sqrt{\sqrt{1+4\Lambda \mathcal{K}}-1}}.
\end{eqnarray}
In standard Electrodynamics, the extremal BH mass corresponds to this expression where the term $\mathcal{K}$ must be replaced just by the sum of squares of the charges, while for 
$\Lambda=0$, this mass simply becomes $\sqrt{\mathcal{K}}$. Since the parameter $\mathcal{K}$ is a monotonically increasing function of $\eta$, it can be seen that for equal parameters the BH extreme mass in the IEM with positive $\eta$ is larger than the extremal BH mass in the standard Electrodynamics theory, whereas in the IEM with negative $\eta$ the extreme BH mass is smaller than the mass of the extremal BH in the standard theory, even disappearing for such a negative $\eta$ that $\mathcal{K}$ becomes negative.
Condition (\ref{condition_extremal}) will be employed in the following sections to discard some sets of parameters in the IEM model.

\begin{figure*}
	\centering
		\includegraphics[width=0.40\textwidth]{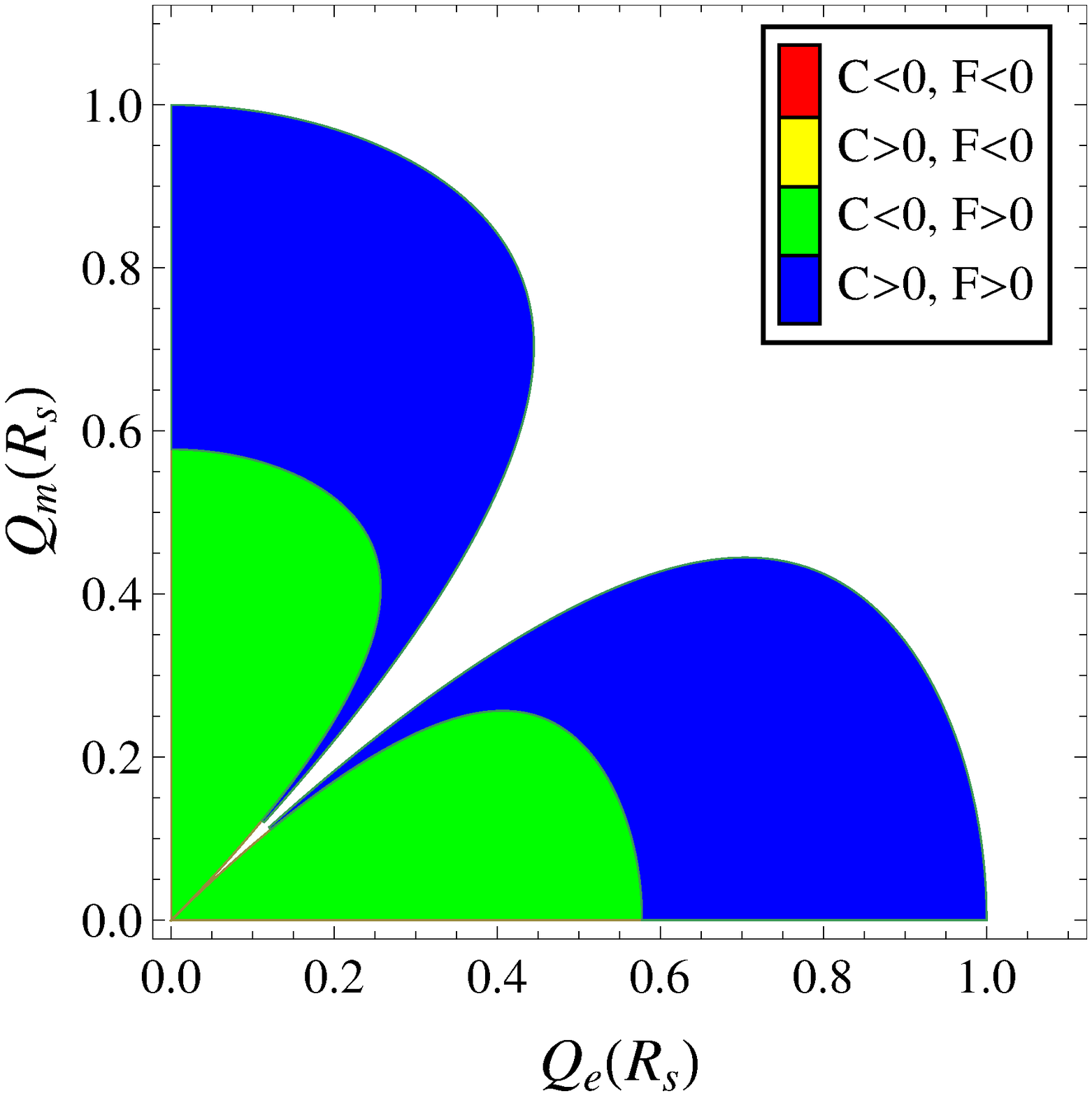}\,\,\,\,\,\,\,\,\,\,\,\,\,\,
		\includegraphics[width=0.40\textwidth]{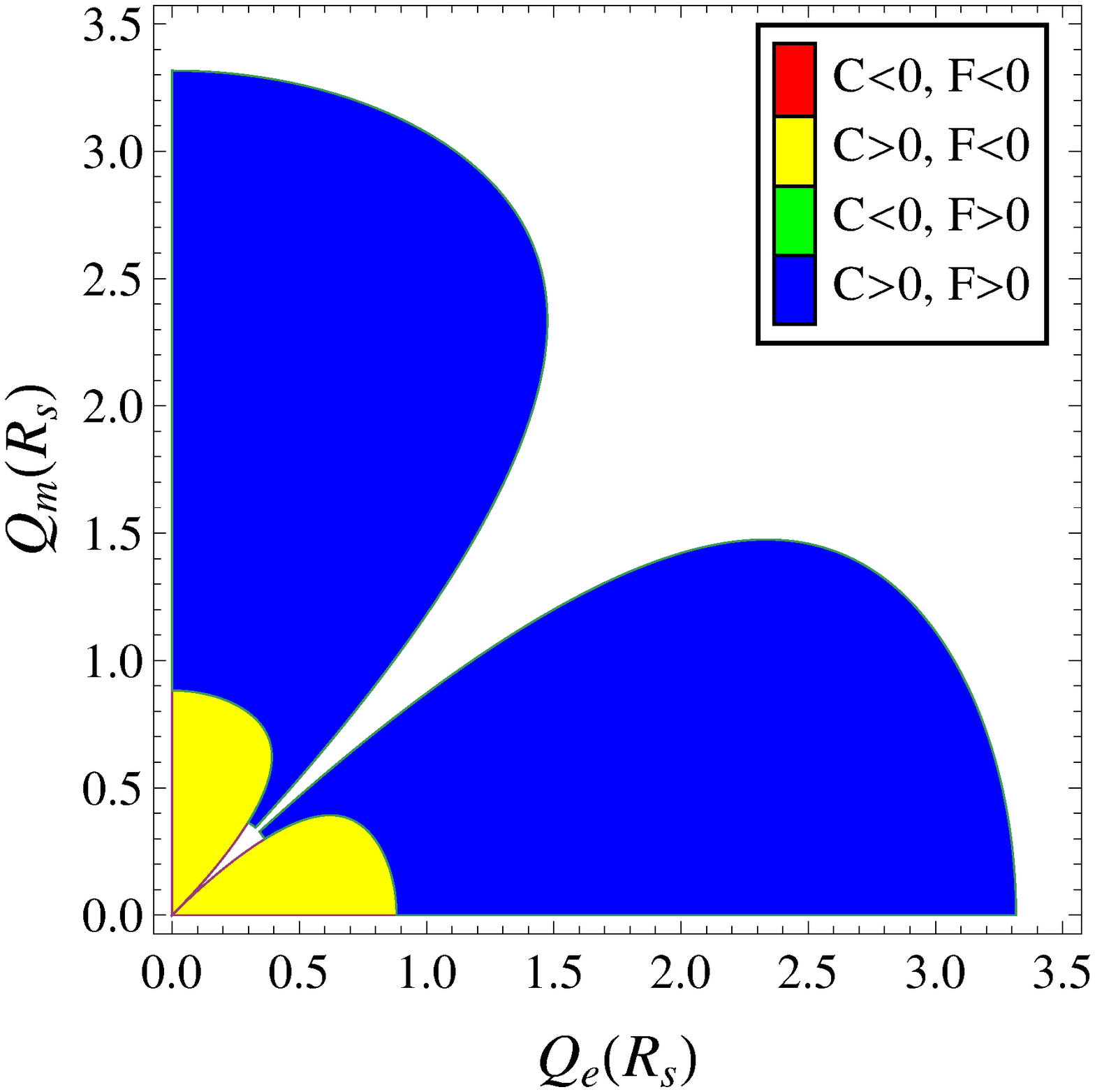}
 \\[1cm] 
		\includegraphics[width=0.40\textwidth]{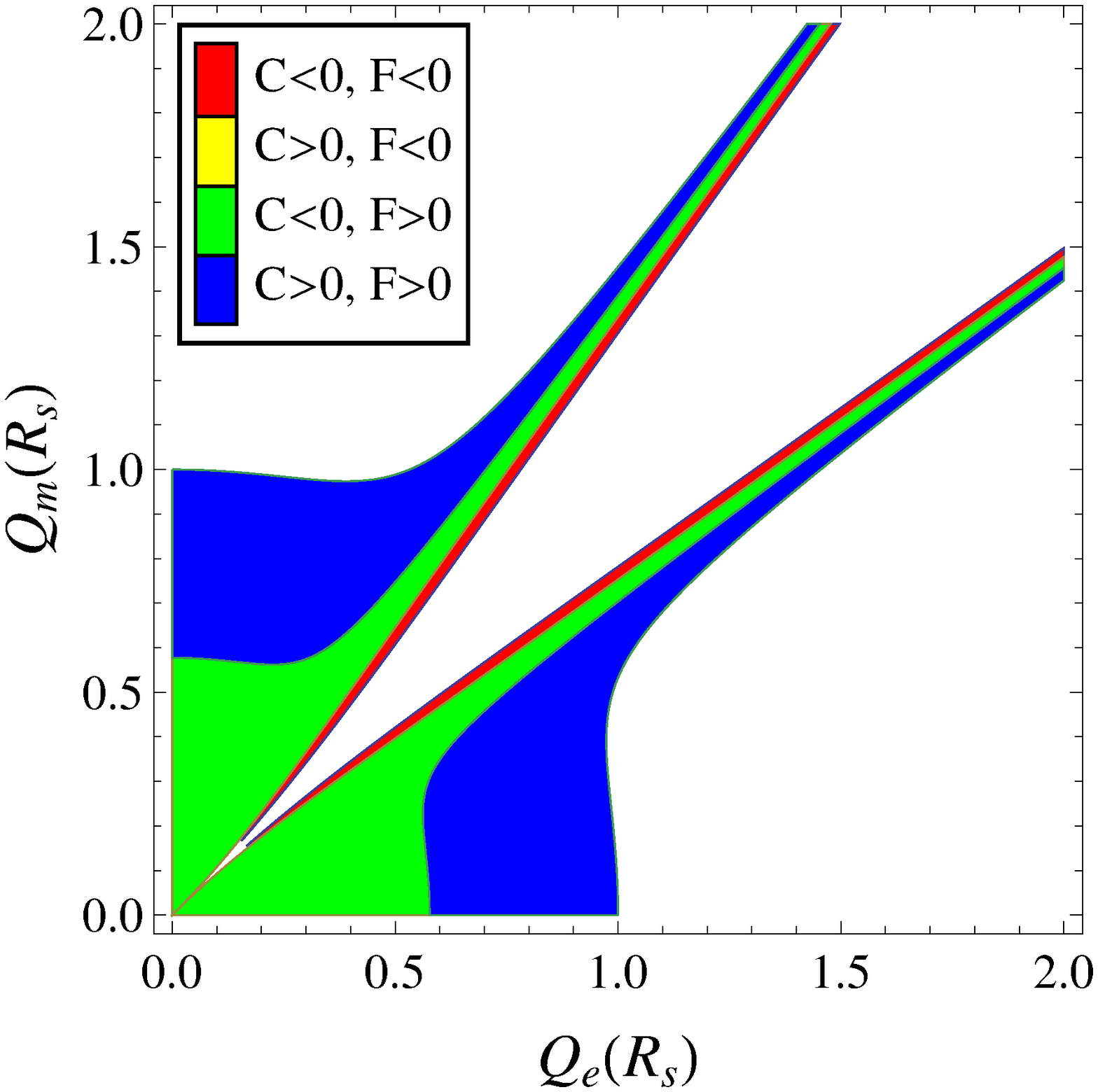}	\,\,\,\,\,\,\,\,\,\,\,\,\,\,
		\includegraphics[width=0.40\textwidth]{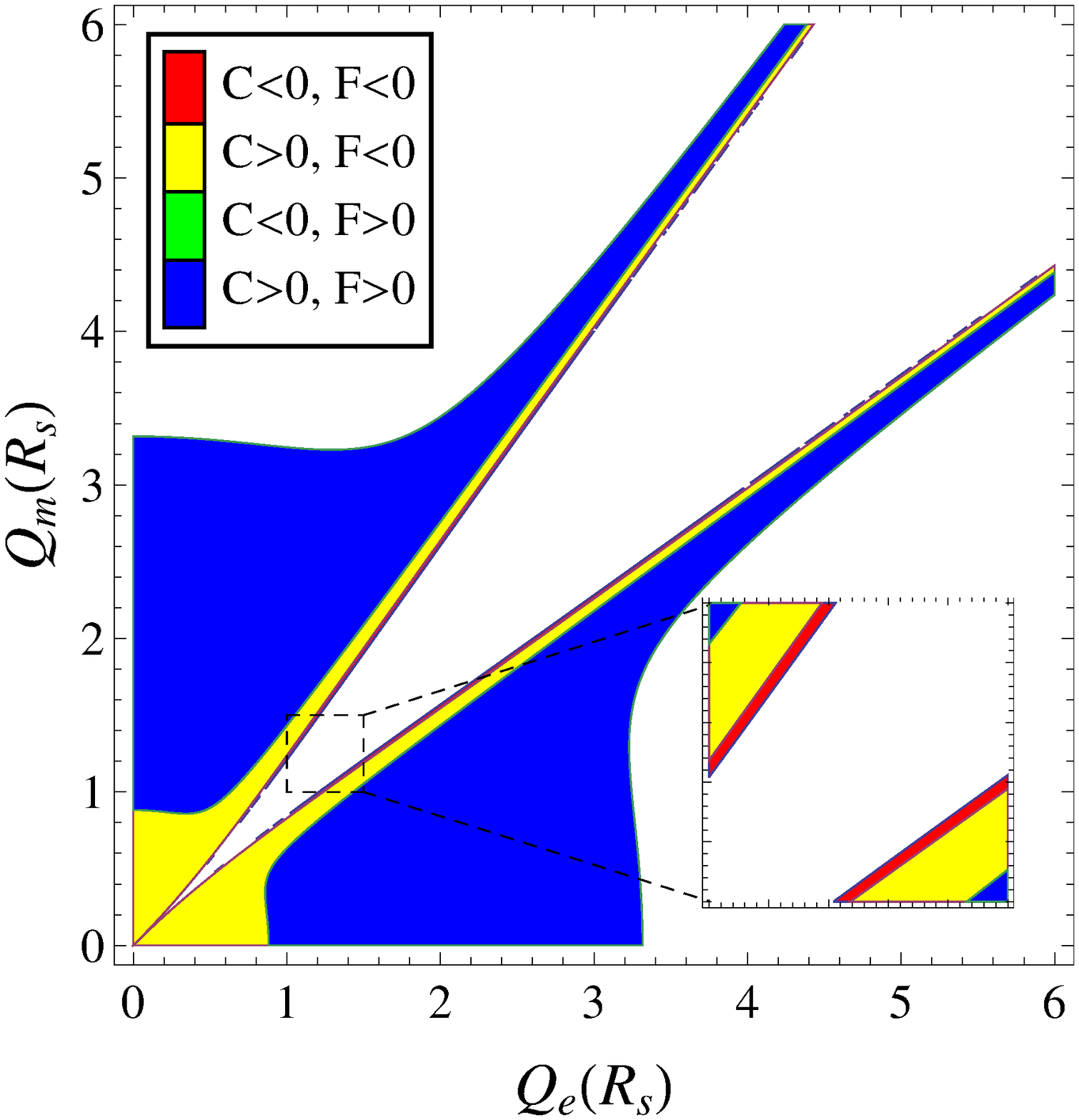}
		\caption{\footnotesize{
		Phase diagrams of BH solutions in the IEM model: The horizon radius was taken $r_{h}=1 R_s$, and different values of $\eta$ and $\Lambda$ are considered. Regions corresponding to super-extremal configurations (those with masses less than the extremal BH mass) or leading to negative masses (if $\mathcal{K}<0$ there is not an extremal BH mass)
were avoided. Phase diagrams for $\eta=0.1$ in flat spacetime ($\Lambda=0$) (upper left panel) and
AdS spacetime with $\Lambda=10\,R_s^{-2}$ with the same value of $\eta$ (right upper panel).
One can appreciate that for positive $\eta$ the phase diagrams are deformed with respect the Standard Electrodynamics, but no new phase is held.  In the lower left and lower right panels, the phase diagrams of the solutions for $\eta=-0.1$, in flat spacetime and in AdS spacetime - with $\Lambda=10\,R_s^{-2}$ - are respectively plotted. Here, we see that for negative $\eta$ a new phase,  where both $C$ and $F$ are negative, appears. Such a phase is not realized in the standard case.
}}
	\label{fig:termo_general}
\end{figure*}

\section{Thermodynamics analysis in $\text{AdS}$ space}
\label{sec:termo}

In this section, we shall apply the so-called Euclidean Action method \cite{Gibbons:1976ue} in order to obtain a thermodynamics analysis of the Reissner-Nordstr\"om-like solution corresponding to the IEM defined by the Lagrangian density \eqref{phimodelo}. We shall focus on the AdS space case ($\Lambda > 0$), in order to avoid
the normalization problem mentioned above. With this method, we shall obtain the thermodynamics properties of the BH solutions. Consequently the BH configurations stability shall thus be studied.

First of all,  the BH temperature can be defined in terms of the horizon gravity $\kappa$ as \cite{Hawking:1974sw}
\begin{eqnarray}
T=\frac{\kappa}{4\pi}\,=\,\frac{1}{4\pi} \lim_{r \rightarrow r_{h}} \frac{\partial_r g_{tt}}{\left| g_{tt} g_{rr}\right|}\, ,
\end{eqnarray}
Replacing \eqref{lambda_general} in the temperature definition, one gets 
\begin{eqnarray}
T=\frac{1}{4\pi r_{h}} \left( 1-\frac{\mathcal{K}}{r_{h}^2} + \Lambda r_{h}^2 \right)\,.
\label{temperatura}
\end{eqnarray}
For large BHs with $r_h \rightarrow \infty$ the temperature goes to infinity, whereas near $r_h \sim 0$ the temperature diverges with its sign opposite to the sign of $\mathcal{K}$. Moreover, let 
remark that the positivity of the temperature \eqref{temperatura} is directly guaranteed by \eqref{condition_extremal}.

Once we have obtained the temperature, we can compute the other thermodynamics quantities.  First we use the Euclidean quantum gravity
definition  \cite{Hartle}
introducing the Euclidean time $t\rightarrow {\rm i}\tau$. When in the total action  \eqref{descomposicion_accion}, we replace the time coordinate by the Euclidean time, the action becomes Euclidean and the metric becomes periodical with a period $\beta$ which coincides with the inverse of the temperature \eqref{temperatura}.
Thus, the Euclidean action reads as
\begin{eqnarray}
\!\!\!\! \Delta S_E = -\frac{1}{16 \pi} \int {\rm d}^4x \sqrt{g } \left[ R-2\Lambda -16 \pi \mathcal{L} \left( X, Y \right)\right]\,.
\label{acciom_euclidea}
\end{eqnarray}
The variation of this action with respect to the metric and electromagnetic tensor yields
\begin{eqnarray}
\!\!\! \delta\Delta S_{E} & = & -\frac{1}{16\pi} \int_{\mathcal{Y}}{\rm d}^{4}x\sqrt{g}\left(G_{\mu\nu}+\Lambda g_{\mu\nu}-8\pi T_{\mu\nu}\right)\delta g^{\mu\nu} \nonumber \\
 & +& 2\int_{\mathcal{Y}}{\rm d}^{4}x\sqrt{g} \nabla_{\mu}\left(\mathcal{L}_{X}F^{\mu\nu}+\mathcal{L}_{Y}F^{*\mu\nu}\right)\delta A_{\nu} \nonumber\\
 & -& 2\int_{\partial\mathcal{Y}}{\rm d}^{3}x\sqrt{h}n_{\mu}\left(\mathcal{L}_{X}F^{\mu\nu}+\mathcal{L}_{Y}F^{*\mu\nu}\right)A_{\nu} \,,
\end{eqnarray}
where $G_{\mu \nu}$ is the Einstein tensor, $n_\mu$ the normal vector to the boundary surface $\partial \mathcal{Y}$,  $h_{\mu \nu}$ the induced metric on $\partial \mathcal{Y}$ and $h$ its determinant. If we impose the bulk terms to be null, we achieve the  Einstein's \eqref{ec_campo} and Maxwell's \eqref{MAX_N} equations.
However since the surface integral has to vanish to have a differentiable action functional and a well defined action principle, it imposes the boundary condition $\delta A_a=0$ on $\partial \mathcal{Y}$, i.e., this action is the appropriate to study the ensemble with fixed electric potential, $A_0$, and fixed magnetic charge. In order to obtain an action valid for an ensemble of fixed constant charges, we have to add to the action a surface term as follows \cite{Caldarelli},
\begin{equation}
\widetilde{\Delta S}_{E}=\Delta S_{E}+2\int_{\partial\mathcal{Y}}{\rm d}^{3}x \sqrt{h}n_{\mu}\left(\mathcal{L}_{X}F^{\mu\nu}+\mathcal{L}_{Y}F^{*\mu\nu}\right)A_{\nu}\,,
\end{equation}
whose variation yields
\begin{eqnarray}
\!\!\!\!\!\!\!\!\!\! \delta\widetilde{\Delta S}_{E} & = & 2 \int_{\partial\mathcal{Y}}{\rm d}^{3}x \sqrt{h} \delta\left[n_{\mu}\left(\mathcal{L}_{X}F^{\mu\nu}+\mathcal{L}_{Y}F^{*\mu\nu}\right)\right] A_{\nu} \nonumber \\
 &+ & \left(\text{bulk terms}\right)\,,
\end{eqnarray}
so now, the vanishing surface term requires the boundary condition $\delta\left(n_{\mu}\mathcal{L}_{X}F^{\mu\nu}+n_{\mu}\mathcal{L}_{Y}F^{*\mu\nu}\right)=0$ at infinity.
The computation of the bulk terms requires its evaluation
as the  four-volume difference of two metrics: the first volume, when there is solely an AdS metric ($M=0$, $Q_e=0$ and $Q_m = 0$); and second one, when there is our metric solution \eqref{lambda_general} ({\it c.f.} Ref.~\cite{Witten:1998zw}). The computation of the difference leads to the expression for the Euclidean action as follows
\begin{eqnarray}
\widetilde{\Delta S}_E & = & \beta\left[-\frac{\Lambda}{12}\left(r_{h}^{3}-\frac{3}{\Lambda}r_{h}\right)+\frac{3}{4}\frac{\mathcal{K}}{r_h}\right]\,.
\label{euclidean_action}
\end{eqnarray}
%
From \eqref{euclidean_action} we can obtain the different thermodynamics quantities. The Helmholtz free energy is just the quotient between the Euclidean action and the inverse of temperature:  $F=\widetilde{ \Delta S }_E / \beta$. Therefore,
\begin{eqnarray}
 F = -\frac{\Lambda}{12}  \left( r_{h}^3-\frac{3}{\Lambda} r_{h} \right) + \frac{3}{4} \frac{\mathcal{K}}{r_{h}}\,.
\label{energia_libre}
\end{eqnarray}
On the other hand, the massive energy is defined as the derivative of the the Euclidean action with respect to the inverse of the temperature,
\begin{eqnarray}
\mathcal{M} =\frac{\partial \widetilde{ \Delta S}_E}{\partial \beta}= \frac{\frac{\partial \widetilde{\Delta S}_E}{\partial r_h}}{ \frac{\partial \beta}{\partial r_h}} = M\,,
\label{Massive_energy}
\end{eqnarray}
As in the standard case, the massive energy is just the mass of the BH which appears in the RN metric.
Additionally, the entropy of the BH is defined as the difference
\begin{eqnarray}
S=\beta \mathcal{M} - \beta F = \pi r_{h}^2\,.
\label{entropia}
\end{eqnarray}
As usual, the achieved BH entropy is just a quarter of the horizon area $A=4\pi r_{h}^2$.  Hence, our result is compatible
with the standard result ensuring the character of the BH entropy as a Noether charge \cite{Wald:1993nt}.

Finally, the heat capacity $C$ can be defined as
$C= T \frac{\partial S}{\partial T}\,,$
so we can replace expressions \eqref{temperatura} and \eqref{entropia} in this definition, yielding
\begin{eqnarray}
C= 2 \pi r_h^2 \frac{\Lambda r_h^4 +r_h^2-\mathcal{K}}{\Lambda r_h^4-r_h^2+3 \mathcal{K}} \,.
\label{capacidad_calorifica}
\end{eqnarray}
Once the relevant thermodynamics quantities are obtained, it is possible to discuss the BH stability regions  in terms of the sign of the
Helmholtz free energy \eqref{energia_libre} and the heat capacity \eqref{capacidad_calorifica} \cite{Hawking:1982dh}. BH configurations with $F>0$ are more energetic than pure radiation, so they eventually decay to radiation by tunneling; whereas BH solutions with $F<0$ will not decay to radiation since they are less energetic. Furthermore, if the solution has $C<0$ it is unstable under acquiring mass, on the contrary to solutions with $C>0$ \cite{Hawking:1982dh}.
%
%
In the following, we discuss the stability regions for the IEM as well as we compare the results with the standard Electrodynamics theory which are briefly revised below.

\begin{figure*}
	\centering
		\includegraphics[width=0.28\textwidth]{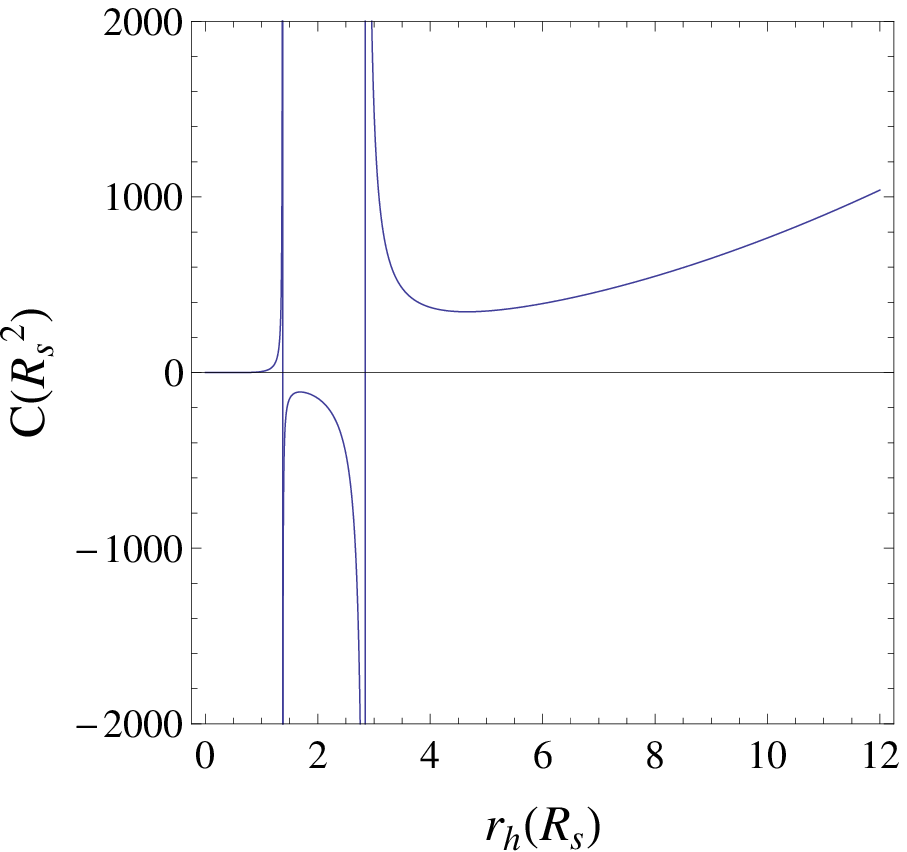}\,\,\,\,\,\,
		\includegraphics[width=0.28\textwidth]{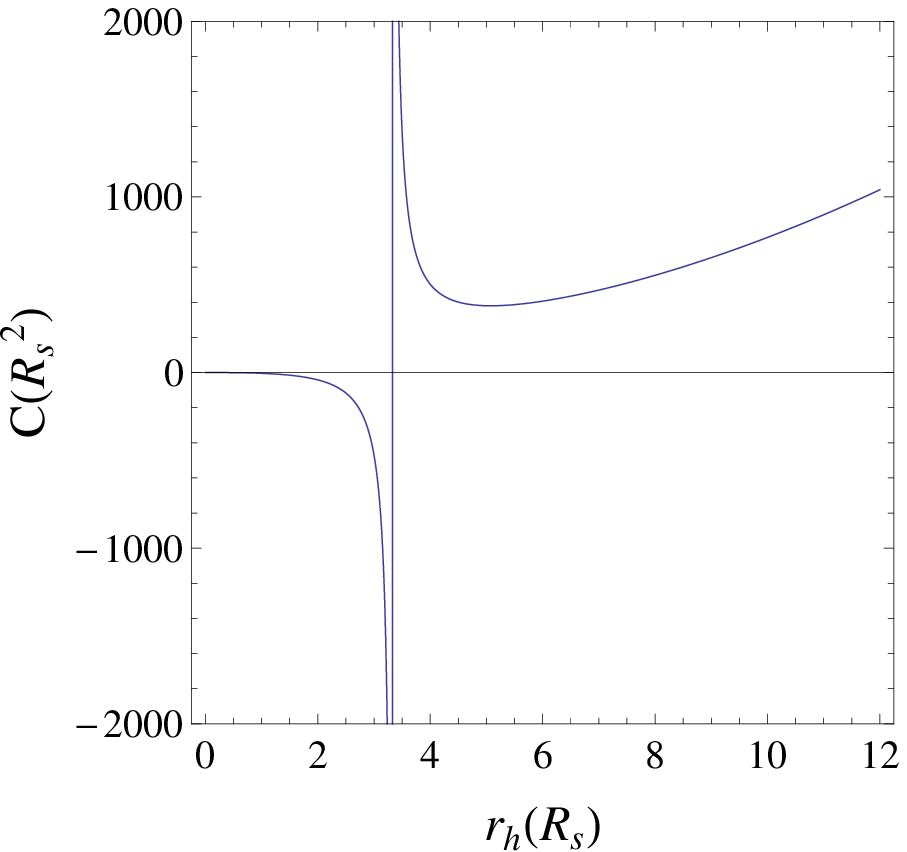}\,\,\,\,\,\,
		\includegraphics[width=0.28\textwidth]{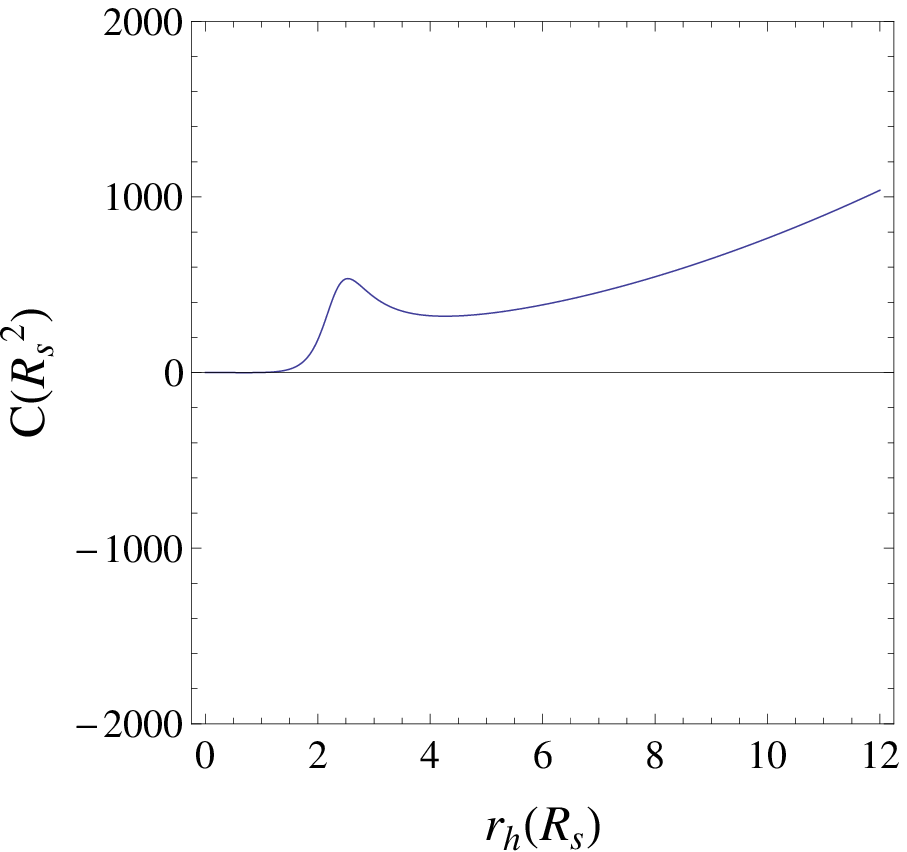}
		\caption{\footnotesize{
		Behavior (from left to right) of the heat capacity for {\it slow}, {\it inverse}  and {\it fast} black holes, respectively. We can see that {\it slow} BHs present two phase transitions (for two horizon radii the heat capacity diverges). {\it Inverse} BHs present a unique phase transition and {\it fast} BHs don't have any phase transitions. For the sake of simplicity, the considered values for each panel were:
		Left: $\eta=-0.1$, $Q_e=0.7 R_s$, $Q_m=0.2 R_s$, $\Lambda=0.1R_s^{-2}$;
		Centre: $\eta=-0.1$, $Q_e=0.5 R_s$, $Q_m=0.4 R_s$, $\Lambda=0.1R_s^{-2}$;
		Right:   $\eta=-0.1$, $Q_e=1 R_s$, $Q_m=0.1 R_s$, $\Lambda=0.1R_s^{-2}$.
}}
	\label{fig:C_speed}
\end{figure*}

\subsection{Standard case: $\eta=0$}
For illustrative purposes let us consider the case $\eta=0$ in the IEM Lagrangian density \eqref{phimodelo}. In this case, $\mathcal{K}$ is just the sum of squares of the charges,
\begin{eqnarray}
\mathcal{K}_{\eta=0}= \left(Q_e^2+Q_m^2\right)\,.
\end{eqnarray}
Using this result, we can simplify the free Helmholtz energy \eqref{energia_libre} and the heat capacity (\ref{capacidad_calorifica}) of the BH solution are given by
\begin{eqnarray}
&&F_{\eta=0}= -\frac{\Lambda}{12} \left( r_{h}^3 -\frac{3}{\Lambda} r_{h}\right) +\frac{3}{4} \frac{Q_e^2+Q_m^2}{r_{h}} \,, \\
&&C_{\eta=0}=2\pi r_{h}^2 \frac{\Lambda r_{h}^4+r_{h}^2- \left(Q_e^2+Q_m^2\right)}{\Lambda r_{h}^4-r_{h}^2+3 \left( Q_e^2+Q_m^2\right)} \,.
\end{eqnarray}
In Figure \ref{fig:termo_estandar}, we represent the phase diagrams of a BH solutions in the Standard Electrodynamics theory in flat and AdS spacetimes. One can see that the phase with both $C$ and $F$ negative does not appear in the standard Electrodynamics theory.

\subsection{General case}

In the IEM, depending on the parameter $\eta$  and the cosmological constant $\Lambda$ the thermodynamics phase corresponding to $\left\{ C<0, F<0\right\}$, which is absent in
the standard Electrodynamics theory, may exist.
In order to illustrate this scenario, we represent
in Figure \ref{fig:termo_general} the phase diagrams for different signs
of the parameter $\eta$ in flat  and AdS  spacetimes.
%
For the example $\eta=0.1$ (positive $\eta$) we see that the phases are deformed with respect to the standard case but new phases do not appear.
%
%
On the contrary, for the example $\eta=-0.1$ (negative $\eta$) we see that both in the flat space and in the  AdS configuration
the new stability phase corresponding to $\left\{ C<0, F<0\right\}$ arises. This means that the BH solutions in the IEM host a different stability phenomenology from the standard Electrodynamics model.

\section{Classification of BH solutions in terms of the number of phase transitions}
\label{sec:termo_speed}

In this section we shall perform a classification of BH solutions based on the number of phase transitions that they present. These phase transitions occur at a set of values of $\Lambda$, $Q_e$, $Q_m$ and $M$ for which the denominator of the heat capacity \eqref{capacidad_calorifica} goes to zero, { i.e.}, the heat capacity goes through an infinite discontinuity \cite{Cembranos:2011sr}. According to the heat capacity definition and by straightforward calculation, that discontinuity turns out to happen whenever the derivative of the temperature \eqref{temperatura} with respect to the external horizon radius is null, i.e., $\left. \frac{\partial T}{\partial r_h} \right|_{\Lambda, Q_e, Q_m}=0$, which leads to the parameters constraint
\begin{eqnarray}
r_h^2 = \frac{1}{2 \Lambda} \left( 1 \pm \sqrt{  1 - 12 \mathcal{K} \Lambda  }  \right) \,.
\label{eq_sec_vel}
\end{eqnarray}
The resolution of the above equation allows us to distinguish three different classes of BH solutions:

\begin{itemize}

\item {\it Fast} BHs. If $\mathcal{K}>\frac{1}{12 \Lambda}$, the radicand in \eqref{eq_sec_vel} is negative and consequently this expression 
is not satisfied for any $r_h$ and therefore phase transitions are absent for these BH configurations . We shall refer to these kinds of solutions as {\it fast} BHs.
In flat spacetime, $\Lambda = 0$, this kind of solution is not allowed.

\item {\it Slow} BHs. For $0< \mathcal{K} <\frac{1}{12 \Lambda}$, equation \eqref{eq_sec_vel} can be satisfied for both plus and minus signs, since for both possibilities $r_h^2>0$. It means that for these BH configurations there are two horizon radii for which a phase transition occurs, i.e., there are two different phase transitions. We shall refer to these kinds of solutions as {\it slow} BHs.

\item {\it Inverse} BHs. Provided $\mathcal{K}<0$, equation \eqref{eq_sec_vel} can be satisfied for the plus sign but not for the minus sign.
In this case, there is solely one  phase transition and we shall refer to these kinds of solutions as {\it inverse} BHs, since they appear in the IEM but not in the standard Electrodynamics theory.
\end{itemize}

\begin{figure*} [htbp]
	\centering
		\includegraphics[width=0.40\textwidth]{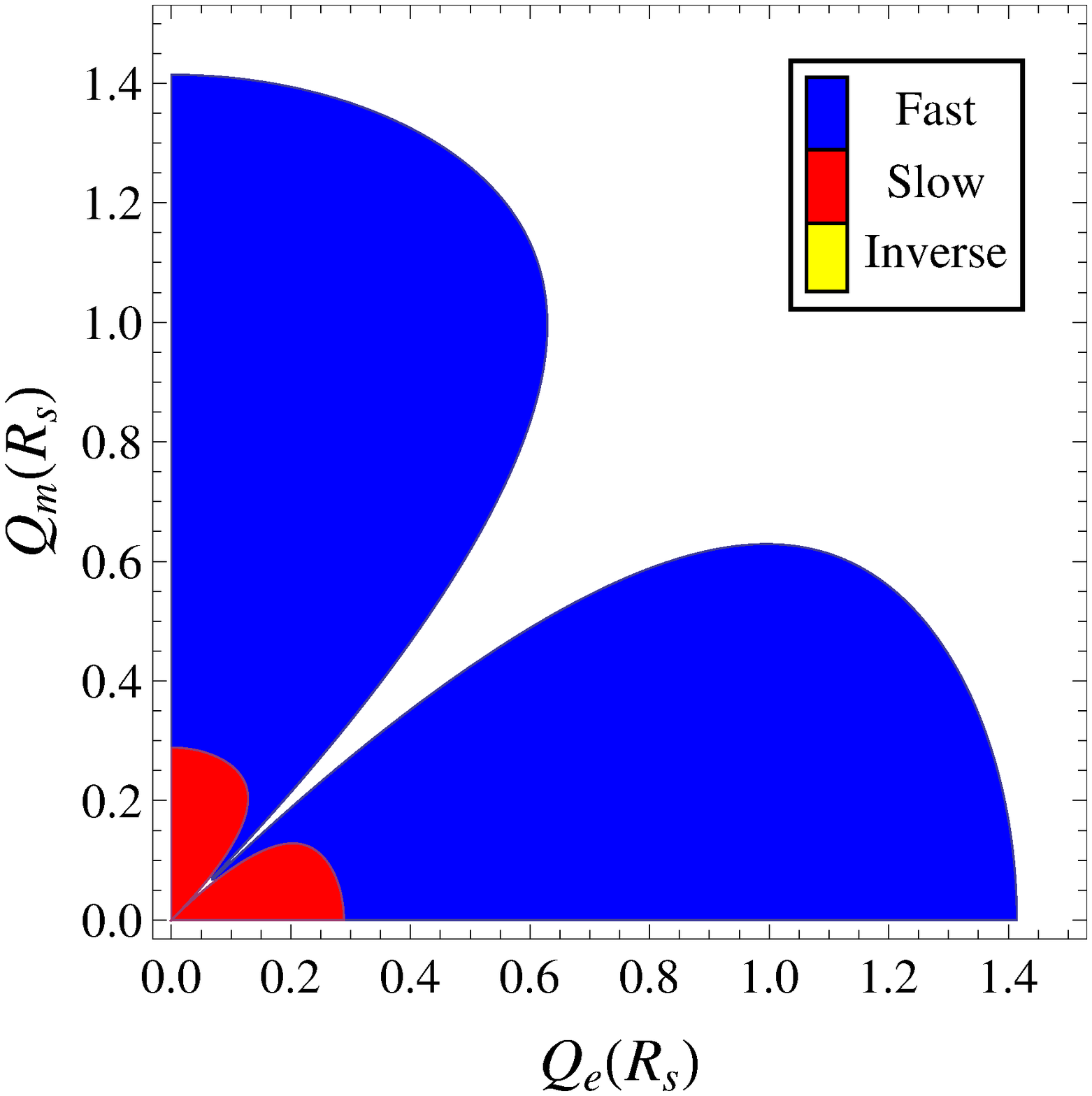}
		\includegraphics[width=0.40\textwidth]{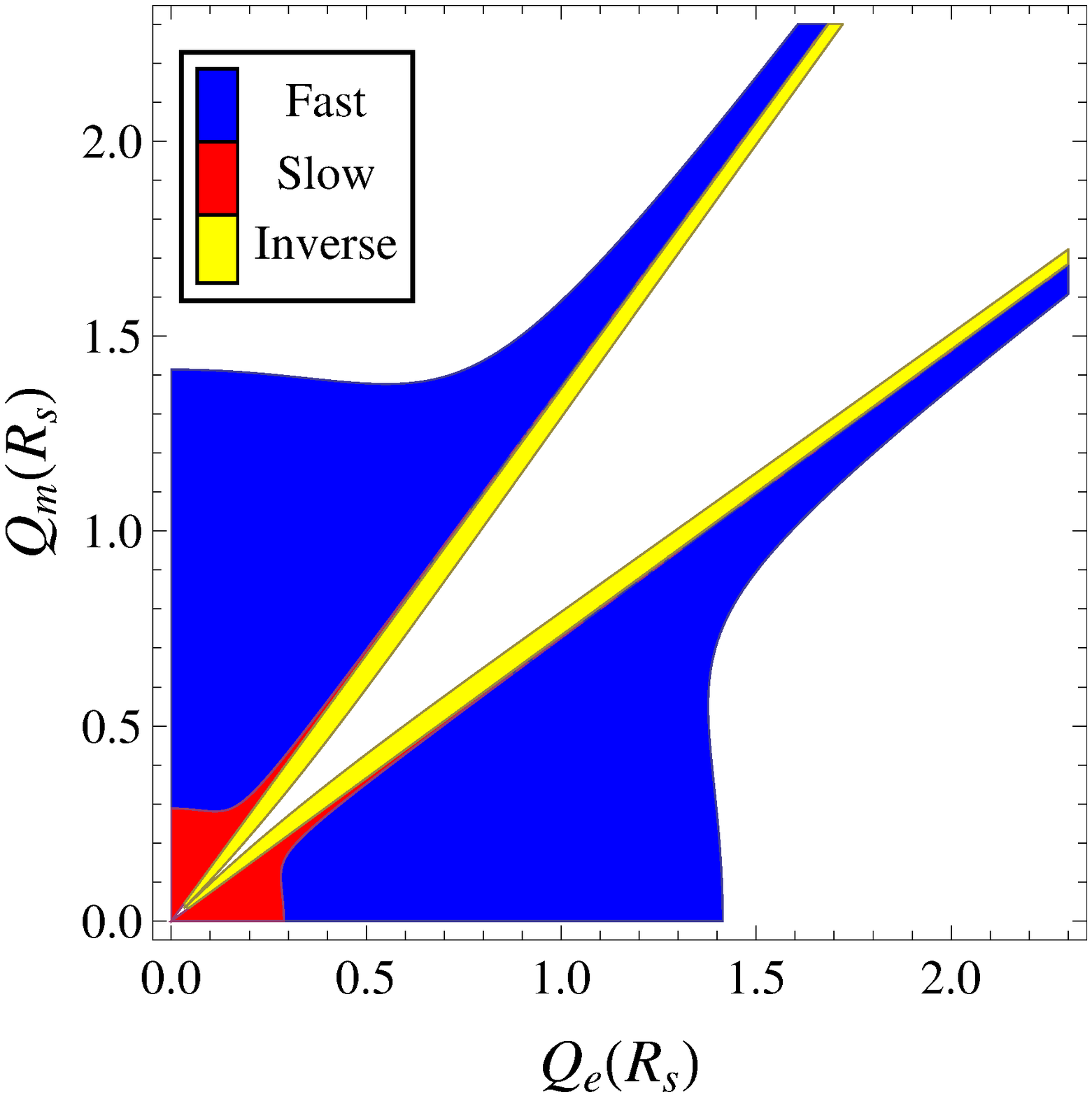}
		\caption{\footnotesize{
		Classification of BHs solutions as a function of the charges $Q_e$ and $Q_m$:  {\it Fast} BHs (blue), {\it slow} BHs in (red) and {\it inverse} BHs (yellow) are depicted.
In the left panel, the case with $\eta=0.1$,  $r_h=1\,R_s$ and $\Lambda= 1\, R_s^{-2}$ is represented, and we can see that the {\it inverse} class does not appear. This is an expected result since for $\eta>0$, the charge term $\mathcal{K}$ is always positive and the {\it inverse} class is not allowed.
In the right panel, regions corresponding to the parameters $\eta=-0.1$,  $r_h\,=\,R_s$ and $\Lambda= 1 \, R_s^{-2}$ are depicted. We can see that the
{\it inverse} BH does appear. In this case, all the classes described in Section \ref{sec:termo_speed} are realized.
		}}
	\label{fig:speed_n2}
\end{figure*}

In Figure \ref{fig:C_speed}, the heat capacity for different classes of BHs is represented. One can distinguish that {\it slow}, {\it inverse} and {\it fast BHs} present two, one or none phase transitions respectively.
On the other hand, in Figure \ref{fig:speed_n2}, we depicted the domain of each class in the case $\eta=\pm 0.1$, $r_h=R_s$ and $\Lambda=1\, R_s^{-2}$. For a negative parameter $\eta$ all the three classes of BHs are present. However, for positive $\eta$, the {\it inverse} type does not appear. This is due to the fact that for positive $\eta$, the charge term given by \eqref{K2(Theta,Qm)} is always positive.

\section{Conclusions}
\label{sec:conclusions}

In this paper we have examined gravitational solutions associated with the Inverse Electrodynamics Model as defined in expression \eqref{phimodelo}. This model, which constitutes a straightforward 
extension of the usual Electrodynamics theory, is parity and gauge invariant, 
and respects conformal invariance.
For a small enough value of the new parameter $\eta$,
the considered model can be interpreted as a perturbation of the standard Electrodynamics theory. 
However,  when the extra contributions are comparable or larger than the standard Maxwell term, 
the qualitative differences between the black-hole  thermodynamics associated with the Inverse Electrodynamics Model  and standard electrodynamics appear.
In this investigation we have precisely focused on the thermodynamical aspects of strongly coupled gravitational systems.
First, we have shown that for static and spherically symmetric U(1) fields,  this model is able to support Reissner-Nordstr\"om-like black-hole solutions.
%
After having obtained the metric tensor, we have performed a thermodynamics analysis of the solutions using the Euclidean Action approach.
For different black-hole types with electrical and/or magnetic charges,  we have thus explored the corresponding phase diagrams of those configurations in the frame of the Inverse Electrodynamics Model.
The stability of those configurations is fully characterized by the signs of the heat capacity and the free Helmholtz energy. We have found that for some sets of values of the Inverse Electrodynamics Model parameters, a new black-hole stability phase, namely a phase where both heat capacity and free energy are negative, which does not appear in the standard Electrodynamics theory, arises.
This phase would imply that the black hole would possess a free energy smaller than pure radiation (null free energy) and consequently
pure radiation will tend to tunnel or to collapse into the black-hole configuration. 
The fact that the heat capacity is negative means that, analogously to the Schwarzschild black holes, the more energy (mass) the black hole acquires the lower its temperature will be. To summarize this configuration would never be in equilibrium with thermal radiation. This fact opens the possibility of further study in other extended electromagnetic theories in order to determine whether this behavior is shared by other
non-linear theories.

Finally, we have classified the black-hole solutions in terms of the existing number of phase transitions, i.e., number of heat capacity divergences
as a function of the horizon radius. Namely
we have described the phenomenology of {\it fast}, {\it slow} and {\it inverse} black holes with none, two and one phase transitions respectively.
This analysis shows explicitly a new difference with respect to the standard Electrodynamics since, whilst in the standard case {\it fast} and {\it slow} black holes are the only existing scenarios, in the Inverse Electrodynamics Model there may also exist a third configuration, the {\it inverse} black hole with a sole phase transition.

\smallskip

{\bf Acknowledgments:}
We would like to thank Pablo Jimeno Romero for his useful advice, and to Luis J.~Garay and Jos\'e Beltr\'an Jim\'enez for helpful discussions.
J.A.R.C. and A.d.l.C.D. acknowledge financial support from MINECO (Spain) projects FPA2011-27853-C02-01,  FIS2011-23000 and Consolider-Ingenio MULTIDARK CSD2009-00064.
J.J. acknowledges financial support the Spanish Ministerio de Educaci\'on, Cultura y Deporte for support through grant FPU-13/02934.
J.A.R.C. thanks
the support of the {\it Becas Complutense del Amo} program.
A.d.l.C.D. thanks Kavli Institute for Theoretical Physics China (KITPC) for their hospitality 
and the ACGC University of Cape Town, for support during the early stages of preparation of this manuscript.
A.d.l.C.D. is also indebted to the Centre de Cosmologie, Physique des Particules et Ph\'enom\'enologie CP3,
Universit\'e catholique de Louvain, Louvain-la-Neuve, Belgium for its assistance with the final steps prior to the release of this manuscript.
J.J. is grateful to the Theoretical Physics and the Atomic, Molecular and Nuclear Physics Departments, Complutense University of Madrid for technical facilities.


\begin{thebibliography}{99}
\addcontentsline{toc}{chapter}{Bibliography}

%
%
%


%
%



\bibitem{Reissner}
  H.~Reissner,
 Ann.\ Phys.\ (Leipz.) {\bf 50} (1916) 106

\bibitem{Nordstrom}
 G.~Nordstr\"{o}m, Proc.\ K.\ Ned.\ Akad.\ Wet.\ {\bf 20} (1918), 1238






\bibitem{Peca:1998cs}
  C.~S.~Peca and J.~Lemos, P.S.,
  Phys.\ Rev.\ D {\bf 59} (1999) 124007
  [gr-qc/9805004].

\bibitem{Barnich:2007uu}
  G.~Barnich and A.~Gomberoff,
  Phys.\ Rev.\ D {\bf 78} (2008) 025025
  [arXiv:0705.0632 [hep-th]].

\bibitem{Jardim:2012se}
  D.~F.~Jardim, M.~E.~Rodrigues and M.~J.~S.~Houndjo,
  Eur.\ Phys.\ J.\ Plus {\bf 127} (2012) 123
  [arXiv:1202.2830 [gr-qc]].


\bibitem{QED} A. Dobado, A. G\'omez-Nicola, A. L. Maroto, and J. R. Pel\'aez, {\it Effective Lagrangians for the Standard Model},  Eds. Springer-Verlag (1997).

\bibitem{Born:1934gh}
  M.~Born and L.~Infeld,
  Proc.\ Roy.\ Soc.\ Lond.\ A {\bf 144} (1934) 425.

\bibitem{BI_others}
  D.~-C.~Zou, S.~-J.~Zhang and B.~Wang,   
  Phys.\ Rev.\ D {\bf 89}, 044002 (2014)   [arXiv:1311.7299 [hep-th]];
  M.~Allahverdizadeh, J.~P.~S.~Lemos and A.~Sheykhi, 
  Phys.\ Rev.\ D {\bf 87}, 084002 (2013)  [arXiv:1302.5079 [gr-qc]];
  S.~Gunasekaran, R.~B.~Mann and D.~Kubiznak,   
  JHEP {\bf 1211}, 110 (2012)   [arXiv:1208.6251 [hep-th]];
  R.~Banerjee and D.~Roychowdhury, 
  Phys.\ Rev.\ D {\bf 85}, 104043 (2012)   [arXiv:1203.0118 [gr-qc]];
  R.~Banerjee and D.~Roychowdhury,   
  Phys.\ Rev.\ D {\bf 85}, 044040 (2012)   [arXiv:1111.0147 [gr-qc]];
  %
  S.~Fernando and D.~Krug, 
  Gen.\ Rel.\ Grav.\  {\bf 35}, 129 (2003)   [hep-th/0306120].

\bibitem{Heisenberg:1935qt}
  W.~Heisenberg and H.~Euler,
  Z.\ Phys.\  {\bf 98} (1936) 714
  [physics/0605038].


\bibitem{Dunne:2012vv}
  G.~V.~Dunne,
  Int.\ J.\ Mod.\ Phys.\ A {\bf 27} (2012) 1260004
   [Int.\ J.\ Mod.\ Phys.\ Conf.\ Ser.\  {\bf 14} (2012) 42]
  [arXiv:1202.1557 [hep-th]].


\bibitem{Fradkin:1985qd}
  E.~S.~Fradkin and A.~A.~Tseytlin,
  Phys.\ Lett.\ B {\bf 163} (1985) 123.

\bibitem{Tseytlin:1997csa}
  A.~A.~Tseytlin,
  Nucl.\ Phys.\ B {\bf 501} (1997) 41
  [hep-th/9701125].


\bibitem{Brecher:1998tv}
  D.~Brecher,
  Phys.\ Lett.\ B {\bf 442} (1998) 117
  [hep-th/9804180].

\bibitem{Euler_others}
  R.~Ruffini, Y.~-B.~Wu and S.~-S.~Xue,   
  Phys.\ Rev.\ D {\bf 88}, 085004 (2013)   [arXiv:1307.4951 [hep-th]].

\bibitem{Other_NonLinear}
  G.~W.~Gibbons and K.~Hashimoto,   
  JHEP {\bf 0009}, 013 (2000)  [hep-th/0007019];
  M.~Hassaine and C.~Martinez,   
  Phys.\ Rev.\ D {\bf 75}, 027502 (2007) [hep-th/0701058];
  M.~Hassaine and C.~Martinez,   
  Class.\ Quant.\ Grav.\  {\bf 25}, 195023 (2008) [arXiv:0803.2946 [hep-th]];
  K.~A.~Bronnikov,   
  Phys.\ Rev.\ D {\bf 63}, 044005 (2001) [gr-qc/0006014];
%
J.~Beltran Jimenez, R.~Durrer, L.~Heisenberg and M.~Thorsrud,   
  JCAP {\bf 1310} (2013) 064   [arXiv:1308.1867 [hep-th]];  J.~Beltran Jimenez, E.~Dio and R.~Durrer, 
  JHEP {\bf 1304} (2013) 030   [arXiv:1211.0441 [hep-th]];
  %
  A.~Burinskii and S.~R.~Hildebrandt,   
  Phys.\ Rev.\ D {\bf 65}, 104017 (2002)   [hep-th/0202066];
  I.~Dymnikova,   
  Class.\ Quant.\ Grav.\  {\bf 21}, 4417 (2004)   [gr-qc/0407072];
  M.~Novello, S.~E.~Perez Bergliaffa and J.~M.~Salim,   
  Class.\ Quant.\ Grav.\  {\bf 17}, 3821 (2000)  [gr-qc/0003052];
  M.~Novello, V.~A.~De Lorenci, J.~M.~Salim and R.~Klippert,   
  Phys.\ Rev.\ D {\bf 61}, 045001 (2000)   [gr-qc/9911085];
%
  G.~J.~Olmo and D.~Rubiera-Garcia,   
  Phys.\ Rev.\ D {\bf 84}, 124059 (2011)   [arXiv:1110.0850 [gr-qc]];
 J.~Beltran Jimenez and A.~L.~Maroto,   
  JCAP {\bf 1012} (2010) 025   [arXiv:1010.4513 [astro-ph.CO]]; 
  Phys.\ Rev.\ D {\bf 83} (2011) 023514   [arXiv:1010.3960 [astro-ph.CO]].

\bibitem{DiazAlonso:2009ak}
  J.~Diaz-Alonso and D.~Rubiera-Garcia,
  Phys.\ Rev.\ D {\bf 81} (2010) 064021   [arXiv:0908.3303 [hep-th]];
  Phys.\ Rev.\ D {\bf 82} (2010) 085024  [arXiv:1008.2710 [hep-th]].


\bibitem{Bardeen:1973gs}
  J.~M.~Bardeen, B.~Carter and S.~W.~Hawking,
  Commun.\ Math.\ Phys.\  {\bf 31} (1973) 161.

\bibitem{Gibbons:1976ue}
  G.~W.~Gibbons and S.~W.~Hawking,
  Phys.\ Rev.\ D {\bf 15} (1977) 2752.



\bibitem{Hawking:1978jz}
  S.~W.~Hawking,
  Phys.\ Rev.\ D {\bf 18} (1978) 1747.

\bibitem{Wald:1984rg}
  R.~M.~Wald,   {\it General Relativity}, (University of Chicago Press, Chicago, U.S.A., 1984).

\bibitem{Hawking:1974rv}
  S.~W.~Hawking, 
  Nature {\bf 248} (1974) 30.


\bibitem{Jackson} J.D.~Jackson, {\it Classical Electrodynamics} Ed: John Wiley \& Sons-3rd ed. (1998).







  \bibitem{Gibbons:1977mu}
    G.~W.~Gibbons and S.~W.~Hawking,
    Phys.\ Rev.\ D {\bf 15} (1977) 2738.

\bibitem{Sahay:2010tx}
  A.~Sahay, T.~Sarkar and G.~Sengupta,
  JHEP {\bf 1007} (2010) 082
  [arXiv:1004.1625 [hep-th]].



\bibitem{Hawking:1974sw}
  S.~W.~Hawking,
  Commun.\ Math.\ Phys.\  {\bf 43} (1975) 199
   [Erratum-ibid.\  {\bf 46} (1976) 206].

\bibitem{Hartle}[ J.B. Hartle and S.W. Hawking, Phys. Rev. D {\bf 13}, 2188 (1976); G.W. Gibbons and M.J. Perry, Proc. R. Soc. London A
358, 467 (1978); G.W. Gibbons and S.W. Hawking, {\it Euclidean Quantum Gravity}, World Scientific, (1993).

\bibitem{Caldarelli}
  M.~M.~Caldarelli, G.~Cognola and D.~Klemm,
  Class.\ Quant.\ Grav.\  {\bf 17} (2000) 399
  [hep-th/9908022].

\bibitem{Witten:1998zw}
  E.~Witten,
  Adv.\ Theor.\ Math.\ Phys.\  {\bf 2} (1998) 505
  [hep-th/9803131].

\bibitem{Wald:1993nt}
  R.~M.~Wald,
  Phys.\ Rev.\ D {\bf 48} (1993) 3427
  [gr-qc/9307038].







\bibitem{Hawking:1982dh}
  S.~W.~Hawking and D.~N.~Page,
  Commun.\ Math.\ Phys.\  {\bf 87} (1983) 577.



\bibitem{Cembranos:2011sr}
  J.~A.~R.~Cembranos, A.~de la Cruz-Dombriz and P.~J.~Romero,
  Int.\ J.\ Geom.\ Meth.\ Mod.\ Phys.\  {\bf 11}, 1450001 (2014)
 [arXiv:1109.4519 [gr-qc]];
%
  AIP Conf.\ Proc.\  {\bf 1458}, 439 (2011)  [arXiv:1202.0853 [gr-qc]].
 





\end{thebibliography}
\end{document}